\documentclass[11pt]{article}
\usepackage{hyperref}
\pdfoutput=1
\begin{document}
\title{Tipstreaming of a drop in simple shear flow in the presence of surfactant}
\author{S. Adami, X.Y. Hu and N.A. Adams \\
\\\vspace{6pt} Institute of Aerodynamics, \\
  Technical University Munich, 85748 Garching, Germany}
\maketitle
\begin{abstract}
We have developed a multi-phase SPH method to simulate arbitrary interfaces containing surface active agents (surfactants) that locally change the properties of the interface, such the surface tension coefficient \cite{Adami2010}. Our method incorporates the effects of surface diffusion, transport of surfactant from/to the bulk phase to/from the interface and diffusion in the bulk phase. Neglecting transport mechanisms, we use this method to study the impact of insoluble surfactants on drop deformation and breakup in simple shear flow and present the results in a fluid dynamics video.

\end{abstract}
\section{Introduction}
Exposing drops to extensional flows such as e.g. a simple Couette flow, the viscous forces along the interface tend to deform the drop and elongate it to an ellipsoid-type shape. The balancing force to stop the deformation due to the shearing is the surface tension. When two pure fluids of different types are in contact, the resulting surface tension force is only proportional to the local curvature and is normal to the interface. Depending on the strength of this force and the viscosity ratio between the two fluids $\lambda$, drops are deformed to a steady ellipsoid shape or break up. The correlation between the breakup behaviour and the flow parameter expressed by the Capillar number $Ca\left(\lambda\right)$ is known as the Grace curve \cite{Grace1982}.\par

Adding surface active agents (surfactants) to a multiphase system can strongly alter the flow phenomena. Surfactants mainly affect the surface tension coefficients between two fluids when replacing fluid molecules at the interface with surfactant molecules. Hence, surface tension gradients along the interface can occur, resulting in Marangoni forces \cite{Scriven1960}.\par

Here, we only focus on the case of insoluble surfactants, i.e. surfactants are initially added to the interface and cannot dissolve to the adjacent fluid phases. Bazhlekov et al. \cite{Bazhlekov2006} studied the effect of insoluble surfactants on drop deformation and breakup in simple shear flow with a boundary-integral method and clearly describe the different breakup modes. However, due to the nature of their method, an interface capturing scheme is required and breakup detection requires special procedures. By the use of a Lagrangian particle method we can avoid these steps and can handle strong interface deformations naturally.\par

\section{Governing equations}
Following the well-known weakly compressible SPH method \cite{Monaghan2005}, we use an equation of state (EOS) to relate the pressure to the density and solve the isothermal Navier-Stokes equations on a moving Lagrangian frame. Formulating the surface force as the gradient of the surface stress tensor \cite{Brackbill1992}
\begin{equation}
    \textbf{F}^{\left(s\right)} = \nabla \cdot \left[ \alpha \left( \mathbf{I} -  
	\mathbf{n}\otimes \mathbf{n}\right)\delta_{\Sigma} \right] 
	= - \left( \alpha \kappa \mathbf{n} +  \nabla_{s} \alpha\right) \delta_{\Sigma} ~,
\end{equation}
surface tension can be split into a normal Capillary force $\alpha \kappa \mathbf{n} \delta_{\Sigma}$ with the curvature $\kappa$, the normal vector of the interface $\mathbf{n}$ and the surface delta function $\delta_{\Sigma}$ and the tangential Marangoni force $\nabla_s \alpha \delta_{\Sigma}$ ($\nabla_{s}$ is the surface gradient operator $\nabla_{s} = \left(\mathbf{I} - \mathbf{n}\otimes \mathbf{n}\right)\nabla$).\par

Assuming insolubility in the phases, the evolution of surfactant on the interface is governed by a diffusion equation 
\begin{equation}
    \frac{d \Gamma}{dt} = \nabla_s \cdot \mathbf{D_s} \nabla_s \Gamma~,
\label{eq:surfactant}
\end{equation}
where $\Gamma$ and $\mathbf{D_s}$ are the interfacial surfactant concentration and the diffusion coefficient matrix (in case of isotropic diffusion $\mathbf{D_s} = D_s\cdot \mathbf{I}$), respectively. 

To close our model, we relate the interfacial surfactant concentration $\Gamma$ to the surface-tension coefficient $\alpha$ by a constitutive equation. Widely used in literature, the Frumkin isotherm or the Langmuir model \cite{Tryggvason2001} are known to agree reasonably well with experimental data. By the purpose of visualization, however, a simple linear relation between $\alpha$ and $\Gamma$ is sufficient and is employed here.

\section{Numerical results}
The results of our simulations are presented in \href{file://anc/movie-tipstreaming-highres.mp4}{Video
1}.\par
As a reference, in the first part of the movie we show the effect of the Capillary number on the droplet behaviour in the shear flow. At sub-critical conditions, the drop deforms to a steady ellipsoid. When the shear forces are strong enough, the surface forces cannot counterbalance the shearing of the drop and breakup occurs.\par

When surfactants are present at the interface, another very complex phenomenon can be observed, namely the so-called tipstreaming. At the tips of the deformed drop the interface is compressed and consequently, the surfactant concentration increases locally. Thus, surface tension gradients develop and Marangoni-forces change the behaviour of the drop dramatically. As the surface tension is very low at the tips of the drop, very sharp tips with a high curvature develop. Under critical conditions, this effect results in a singularity where the tipstreaming starts. In our simulations we can show that the thickness of the stream is of the order of the size of a particle.\par

At the end, we show the possibilities to influence the tipstreaming by changing the strength of the surface diffusion. When diffusion is strong, concentration gradients are smoothed and the Marangoni-forces are weaker. Hence, the curvature at the tips can be reduced and tipstreaming is suppressed. Another way of controlling the tipstreaming is by changing the properties of the surfactant, which is not shown here.
 
\section*{Acknowledgment}

We want to acknowledge the support of J. Biddiscombe \cite{Biddiscombe2008} with \textit{pv-meshless}, which was used to visualize the simulation data. Furthermore, we are grateful to R. Fraedrich, who provided the volume rendering of our results \cite{Fraedrich2010}.

\bibliographystyle{plain}
\bibliography{bibdata}

\end{document}